\documentclass[prb,aps,amssymb,shownopacs,twocolumn,superscriptaddress,times,10pt]{revtex4-1}
\usepackage{amsmath}
\usepackage{amssymb}
\usepackage{amsfonts}
\usepackage{listings}
\usepackage{enumerate}
\usepackage{latexsym}
\usepackage{bm}
\usepackage{graphicx}
\usepackage{braket}
\usepackage[pdftex,colorlinks=false]{hyperref}
\usepackage{xcolor}
\usepackage{verbatim}
\usepackage{float}
\usepackage{multirow}

\usepackage{times}

\setcitestyle{numbers,square}

\begin{document}

\tolerance 10000

\newcommand{\cbl}[1]{\color{blue} #1 \color{black}}

\newcommand{\vk}{{\bf k}}

\widowpenalty10000
\clubpenalty10000

\title{Neural Network-based Classification of Crystal Symmetries from X-Ray Diffraction Patterns}

\author{Pascal Marc Vecsei}
\address{
 Department of Physics, University of Zurich, Winterthurerstrasse 190, 8057 Zurich, Switzerland
}

\author{Kenny Choo}
\address{
Department of Physics, University of Zurich, Winterthurerstrasse 190, 8057 Zurich, Switzerland
}

\author{Johan Chang}
\address{
Department of Physics, University of Zurich, Winterthurerstrasse 190, 8057 Zurich, Switzerland
}

\author{Titus Neupert}
\address{
 Department of Physics, University of Zurich, Winterthurerstrasse 190, 8057 Zurich, Switzerland
}

\begin{abstract}
Machine learning algorithms based on artificial neural networks have proven very useful for a variety of classification problems. Here we apply them to a well-known problem in crystallography, namely the classification of X-ray diffraction patterns (XRD) of inorganic powder specimens by the respective crystal system and space group. Over 10$^5$ 
theoretically computed powder XRD patterns were obtained from inorganic crystal structure databases and used to train a deep dense neural network. For space group classification, we obtain an accuracy of around 54\% on experimental data. Finally, we introduce a scheme where the network has the option to refuse the classification of XRD patterns that would be classified with a large uncertainty. This enhances the accuracy on experimental data to 82\% at the expense of having half of the experimental data unclassified. With further improvements of neural network architecture and experimental data availability, machine learning constitutes a promising complement to classical structure determination methodology.
\end{abstract}

\date{\today}

\maketitle

\section{Introduction}
Elastic scattering of plane waves is a powerful technique to reveal the structure of translationally invariant systems.
Diffraction has been successfully applied to atomic~\cite{Rietveld1969}, magnetic~\cite{RODRIGUEZCARVAJAL1993}, 
superconducting vortex~\cite{Eskildsen2011} and even protein 
lattices~\cite{KendrewNature1958} to uncover their crystal structures. Such structure determination can have far reaching implications and 
prominent examples are: 
(i) The correct helical DNA structure was first inferred from interpretation of diffraction patterns~\cite{Frankling1953,Watson1953}.
(ii) For the production of widely used carbon steel avoiding certain crystal structures by thermal quenching is imperative for 
high-quality material properties~\cite{Chipman1972}. (iii) In drug design, diffraction determination of protein structures is crucial~\cite{Song}. 
The determination of crystal structures based on diffraction experiments is therefore an important activity. A sub-problem is the determination of 
crystal symmetries.
As there is a finite number of space groups, 
interpretation of diffraction patterns is then in essence a classification problem. 
A prevalent strategy is to model the diffracted Bragg reflections through the form and structure factors~\cite{Rietveld1969}. 
Crystal structure determination is then settled by the best fit to the experimental data. 
The recent broad dissemination of machine learning algorithms in science, game theory and technology is driven, for a large part, 
by the ability of neural networks to -- after training -- classify data~\cite{stanev2018machine,zhang2018using,pilania2016machine,isayev2017universal,seko2014machine,balachandran2015materials,carleo2017solving,ch2017machine,van2017learning,carrasquilla2017machine,torlai2017neural,wigley2016fast,baldi2014searching,Schindler2017,ParkNNXRD2017,SilverSCIENCE2018, Ziletti2018}.
There, neural networks are able to acquire implicit knowledge through which classification can be achieved without 
prior information about which features in the data are relevant for the classification. It is thus of great interest to apply 
these algorithms to the interpretation of diffraction patterns~\cite{ParkNNXRD2017}.
 
In this work, we study the problem of space group determination from powder X-ray diffraction (XRD) patterns using artificial neural networks. We restrict ourselves to inorganic nonmagnetic materials and employ a fully supervised learning scheme for this classification task.
To train the network, we generate a large amount of data theoretically, by computing diffraction patterns based on crystal structure information of real crystals obtained from various databases \cite{bergerhoff1987crystallographic,Downs2003,Grazulis2009,Grazulis2012,Merkys2016,Grazulis2015,Jain2013}. The trained network is then tested on both theoretical and experimental data, the latter being obtained from the RRUFF database \cite{RRUFF}. 
A recent study~\cite{ParkNNXRD2017} reported, using a convolutional neural network architecture, a remarkable classification accuracy of over 80\% on theoretical data despite failing to correctly classify the few experimental data they obtained. 
We contrast the performance of this convolutional network with a simple dense network and demonstrate that the dense network performs significantly better on experimental data. Furthermore, it is found that when the network misclassifies a structure, the wrongly predicted space group often differs from the correct one by only a few symmetry elements.
Finally, we show that a classification accuracy of above 80\% can be achieved for experimental data, if about half of the data is left unclassified because the network is not certain about it. This is practically relevant since it is better to know that a network is uncertain than for the network to give a wrong classification.

Our results demonstrate that neural networks can be useful in classifying experimental XRD patterns by the space group of the crystal even when training is done only with theoretically computed data. While the accuracy achieved in this classification is not sufficient to rely on this method alone, it may be used to enhance and complement existing algorithms. In any case, the classification accuracy of 50\% we obtain for experimental data is already a considerable achievement given that (i) there are 230 space groups to differentiate, (ii) the experimental data is not perfect due to finite counting statistics and backgrounds, crystal defects, and minority phases, (iii) no prior knowledge about the task is used in training the network and the data is not pre-processed. 

The paper is structured as follows: In Sec.~\ref{sec: problem} we explain in more detail the problem that the neural network is set out to solve. Section~\ref{sec: NN} introduces the structure of the neural networks that we use as well as the training algorithm. Section~\ref{sec: data} shows how the theoretical training data is prepared, and Sec.~\ref{sec: results} discusses all the results including the performance comparison between the convolutional and the dense network.

\section{X-ray Diffraction and Crystal Symmetries}
\label{sec: problem}
\begin{figure}[h]
            \includegraphics[width=0.4\textwidth]{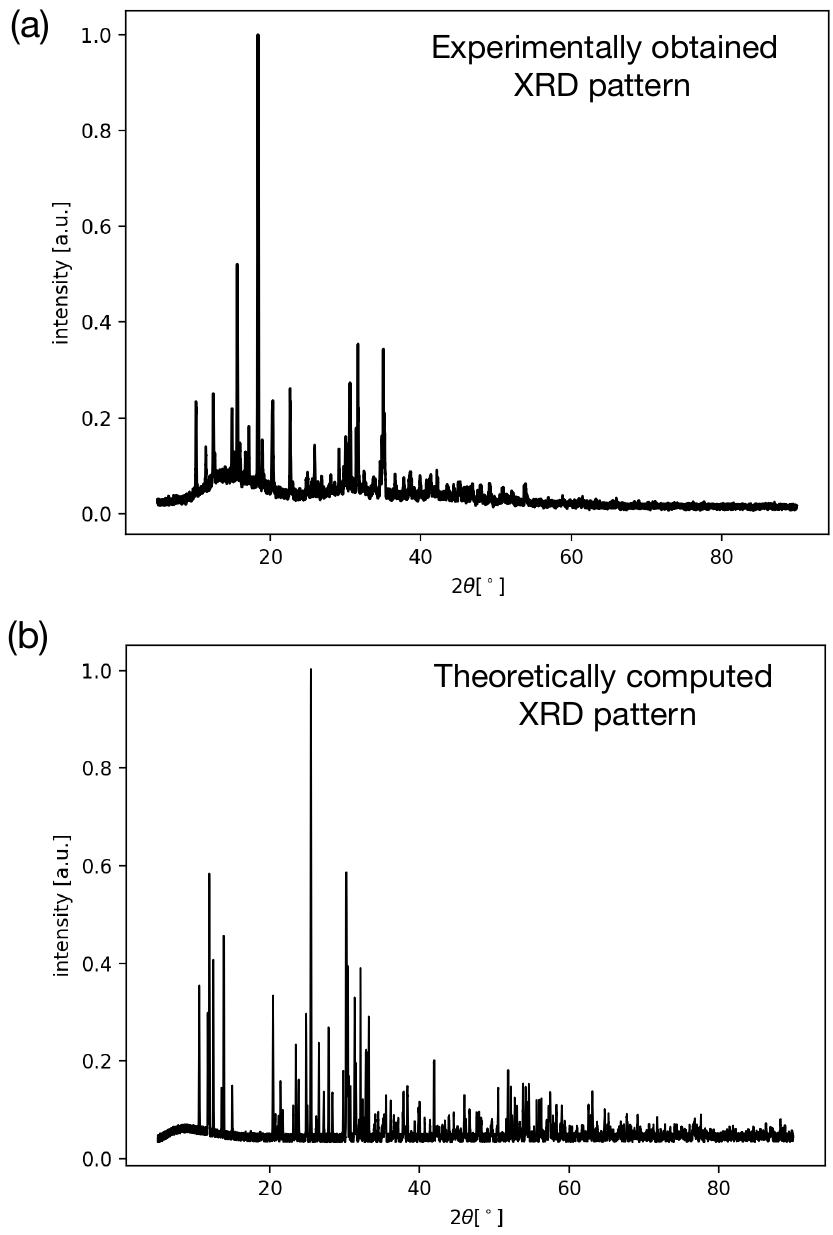}
    \caption{Typical samples of powder XRD patterns. (a) XRD pattern for Borax $ \text{Na}_2 \lbrack \text{B}_4\text{O}_5\text{(OH)}_4 \rbrack \cdot 8\text{H}_2\text{O}$  (obtained from the RRUFF database); (b) Theoretically computed XRD pattern for $\text{C}_{2} (\text{Si}_6 \text{Cl}_{14} )$ including noise and background. Both compounds belong to the space group 15.}
\label{fig: XRD Sample}
\end{figure}

The goal of our machine learning task is to predict 
crystal symmetry (crystal system and space group) from powder XRD patterns. There are
seven crystal systems and $230$ space groups.
In an XRD experiment, 
the photons scatter off the atoms in the crystal, interfere and produce Bragg diffraction patterns
which are collected by a detector. Since the sample is powdered, all crystal orientations are represented and hence the diffraction pattern shows rings centred about the beam axis, i.e., the intensity pattern $S(2\theta)$ is a function of the scattering angle $2\theta$ with respect to the beam axis.
In Fig.~\ref{fig: XRD Sample}, examples of such diffraction patterns are shown. Since the interference depends on the relative positions of the atoms within the crystal, information regarding the crystal symmetries is contained within the XRD pattern. 

Conventional methods \cite{werner1985treor,visser1969fully,boultif1991indexing,le2004monte,altomare2009expo2009,neumann2003x} for obtaining the space group from the XRD pattern generally involve performing some peak finding algorithm. 
Our objective is to train a neural network such that it learns the relevant features of the interference pattern in order to correctly predict the crystal symmetry without relying on preprocessing of the data to identify peaks.
The XRD pattern, i.e., the Bragg peak positions in $2\theta$, depends on the wavelength of the photons. Generally, the patterns can be expressed in terms of scattering wavevectors $q = 4\pi \sin(\theta)/\lambda$. However, for direct comparison with Ref.\cite{ParkNNXRD2017}, we fixed the wavelength at the copper K$\alpha$-line ($\lambda = 1.54$\AA) and hence we  
display our results as a function of $2\theta$.
Since a neural network takes a finite number of inputs, we express the intensity function $S(2\theta)$ as a vector. We set the range from $5^\circ<2\theta<90^\circ$ with a spacing of $0.01^\circ$. In addition, we normalize the functions such that its largest value is one.

\section{Neural Network and Training Algorithm}
\label{sec: NN}
Artificial neural networks are variational approximations to arbitrary functions. Many variants and architectures exist. Here, we consider the so-called feedforward neural networks (FFNN). They consist of a series of successively applied maps. Each map constitutes a ``layer'' of the network. By definition, the first layer is the input layer and the last layer is the output layer of the network. 

Let $\boldsymbol{v}_n$ be the output of layer $n$ and define the input to the network to be $\boldsymbol{v}_0$. The output of layer $n$ is then the input for layer $n+1$. At each layer, we perform a transformation to go from one layer to the next. The most important layer consists of  an affine map followed by a non-linear function (or activation function) $g_n$, i.e., 
\begin{equation}
\boldsymbol{v}_n \rightarrow \boldsymbol{v}_{n+1} = g_n(\boldsymbol{W}_{n}\boldsymbol{v}_{n} + \boldsymbol{b}_{n}),
\end{equation}
where $\boldsymbol{W}_{n}$ is the so-called weight matrix and  $\boldsymbol{b}_{n}$ is the bias vector. When no further constraints are imposed on the weight matrix and bias vector, the layer is called a dense layer. In a so-called convolutional layer, by contrast, not all elements of the weight matrix can be freely chosen. There is a set of constraints which reduces the number of learnable parameters. This allows one to effectively use networks for much larger dimensions of the input vectors and to build both ``deeper''  and ``wider'' networks consisting of more layers. There are other types of layers with specific purpose, such as pooling layers and dropout layers~\cite{Goodfellow-et-al-2016}.

An important freedom in the setup of the network is the choice of activation function $g_n$. For all layers apart from the final output layer, we use the rectified linear unit (ReLU) defined by
\begin{equation}
\textrm{ReLU}(z) = \textrm{max}(0,z)\, ,
\end{equation}
which is applied element wise to the vector $\boldsymbol{v}_n$.  For the output layer, we use the softmax function given by
\begin{equation}
\textrm{SoftMax}(\boldsymbol{v}_n)_{i} = \frac{\textrm{exp}(v_{n,i})}{\sum_{j} \textrm{exp}(v_{n,j})}.
\end{equation}
This softmax function is positive definite and normalized to 1 such that the output layer can be interpreted as a probability distribution. 

\begin{figure*}[t]
            \includegraphics[width=0.9\textwidth]{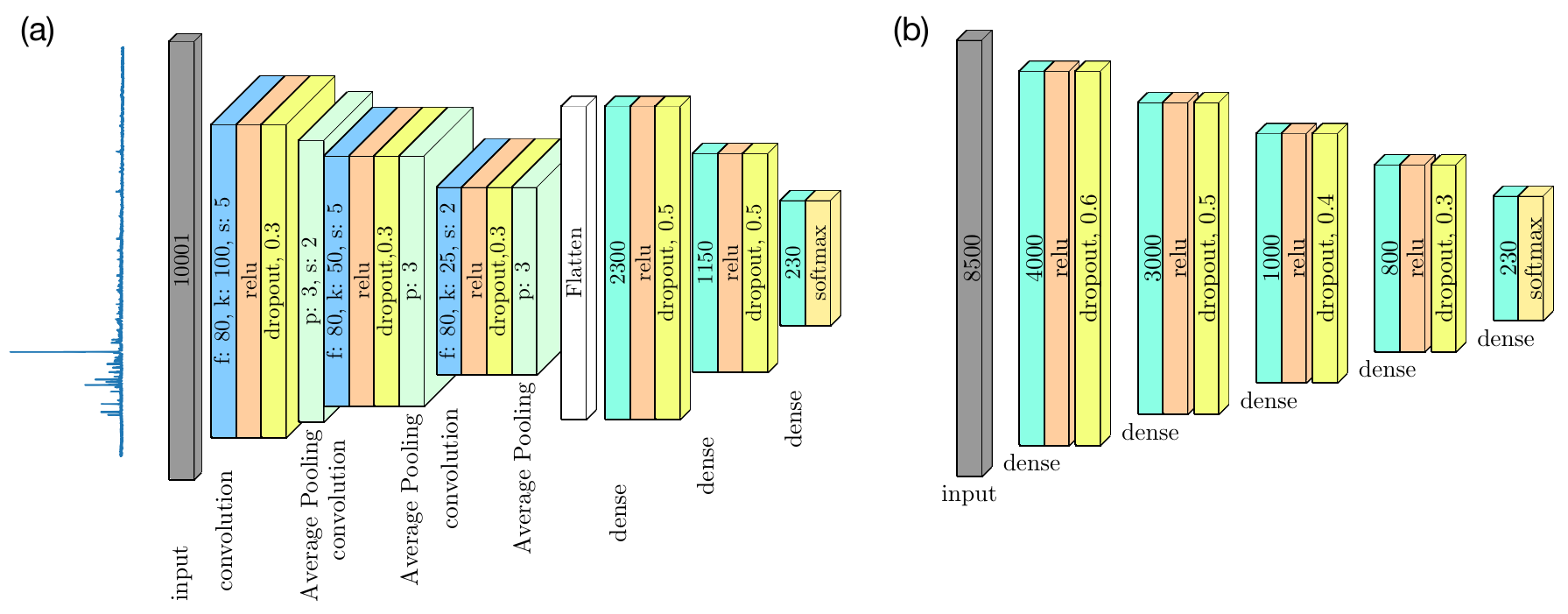}
    \caption{Network architectures used for classifying space groups from powder XRD patterns. (a) Convolutional neural network as used in Ref.~\cite{ParkNNXRD2017}; (b) Dense network. Blue layers represents convolutional layers with $f$ filters of size $k$ and stride $s$, green layers correspond to dense layers with the number of nodes given by the value within and yellow layers are dropout layers with the indicated dropout rate. The gray layers are the input layers with the number denoting the corresponding input size. The XRD patterns are zero padded to match the input size. Evaluated on the experimental samples from the RRUFF database, we achieved a accuracy of $42\%$ for the convolutional network and an accuracy of $54\%$ for the dense network.}
\label{fig: Network Architecture}
\end{figure*}

In this paper, we consider two different network architectures, a convolutional network (this is the same architecture as in Ref.~\cite{ParkNNXRD2017}) and a dense network. We use a one-hot encoding for representing the different possible classification categories. This means that the final layer of the network has as many nodes as there are categories  for a certain classification of the input data, i.e., $7$ nodes if we are classifying crystal systems and $230$ if we are classifying space groups.
A fully confident classification into category $i$ then corresponds to an output vector equal to the unit vector in $i$ direction.  More generally, the network's prediction is taken to be the category with the largest component of the output vector.
 The precise network architectures we used for space group classification are shown in Fig.~\ref{fig: Network Architecture}. Notice that the final layer contains $230$ nodes corresponding to different possible space groups. The networks used for classifying crystal systems are nearly identical apart from the final few layers since the output layer should have only $7$ nodes.

We now proceed to explain how the network is trained. 
For that, we use a labelled dataset $\lbrace (\boldsymbol{x}, l_{\boldsymbol{x}}) \rbrace$ where $\boldsymbol{x}$ is the sample input (the XRD pattern) and $l_{\boldsymbol{x}}$ is the correct label (the space group or crystal system to which the pattern belongs). Let the network output distribution corresponding to the input $\boldsymbol{x}$ be given by $\boldsymbol{y}_{\boldsymbol{x}}$, and that corresponding to the correct label is defined via its components  $y^{c}_{\boldsymbol{x},i} = \delta_{i,l_{\boldsymbol{x}}}$. The network is trained by minimizing the following cost functions:
 the mean-squared-error or quadratic cost
 \begin{equation} 
 \mathcal{C}_{\textrm{quad}} = \frac{1}{n} \sum_{\boldsymbol{x}} \left| \boldsymbol{y}_{\boldsymbol{x}} - \boldsymbol{y}^{c}_{\boldsymbol{x}} \right|^2 
 \end{equation}
 and the categorical cross entropy
 \begin{equation} 
 \mathcal{C}_{\textrm{cross}} = - \frac{1}{n} \sum_{\boldsymbol{x}} \left[ \boldsymbol{y}^{c}_{\boldsymbol{x}} \cdot \ln \boldsymbol{y}_{\boldsymbol{x}}+ (\boldsymbol{1}- \boldsymbol{y}^{c}_{\boldsymbol{x}} ) \cdot  \ln(\boldsymbol{1}-\boldsymbol{y}_{\boldsymbol{x}}) \right],
 \end{equation} 
where the sum is over the training dataset. To minimize the cost function, we use a gradient decent type optimizer called adaptive moment estimation (Adam) \cite{Adam} which proceeds as follows. We denote by $\alpha_r$ the collection of all network parameters, i.e., all the entries of weight matrices and bias vectors, where $r$ indexes all these quantities.
\begin{enumerate}
\item At step $t$, compute the gradient of the cost function $\mathcal{C}$ with respect to the network parameters,
\begin{equation}
g_{r;t} = \nabla_{{\alpha}_r} \mathcal{C}
\end{equation}
\item Compute the decaying first and second moments of the past gradients
\begin{equation}
\begin{split}
m_{r;t} &= \beta_{1} m_{r;t-1} + (1-\beta_{1})g_{r;t} ,\\
v_{r;t} &= \beta_{2} v_{r;t-1} + (1-\beta_{2})g_{r;t}^2,
\end{split}
\end{equation}
where $\beta_1$ and $\beta_2$ are the hyperparameters controlling the decay rate.
\item Because these moments are initialized to zero ($m_{r;0} = v_{r;0} = 0$), there is a bias towards zero. To counteract this, we define the bias-corrected moments 
\begin{equation}
\begin{split}
\hat{m}_{r;t} &= \frac{m_{r;t}}{1 - \beta_{1}^{t}}, \\
\hat{v}_{r;t} &= \frac{v_{r;t}}{1 - \beta_{2}^{t}} .
\end{split}
\end{equation}
\item Update the parameters of the network as
\begin{equation}
\alpha_r \rightarrow \alpha_r - \frac{\eta}{\sqrt{\hat{v}_{r;t}} + \epsilon} \hat{m}_{r;t} ,
\end{equation}
where $\eta$ is the learning rate and $\epsilon$ is a smoothing term to prevent division by zero.
\end{enumerate}
We used the standard values for the hyperparameters of the Adam optimizer: $\eta = 0.001$, $\beta_{1} = 0.9$, $\beta_{2} = 0.999$ and $\epsilon = 10^{-8}$.

In order to speed up the training, instead of summing over the entire dataset in the cost functions $\mathcal{C}$, we can perform the sum over a randomly selected batch of samples. This introduces noise in the computation of the derivative. The smaller the batch size, the larger the noise. It is thus necessary to find a trade-off between speed and accuracy. However, it should be noted that a certain amount of noise can be useful to prevent the network from getting trapped in local minima. 

In each iteration, a batch of samples from the training data is fed to the network and the parameters are updated according to the algorithm above. An epoch is the number of iterations needed to transverse the entire dataset. During the training, we alternate between the quadratic cost $\mathcal{C}_{\textrm{quad}}$ on even epochs and the cross entropy $\mathcal{C}_{\textrm{cross}}$ on odd epochs. The reason for this is that the categorical cross entropy helps to speed up convergence while the quadratic error is necessary for the network to produce a more meaningful output distribution. Without the quadratic error, the output of the dense network is always close to a one-hot encoded vector even for wrong predictions. 

\begin{figure*}[t]
            \includegraphics[width=1.0\textwidth]{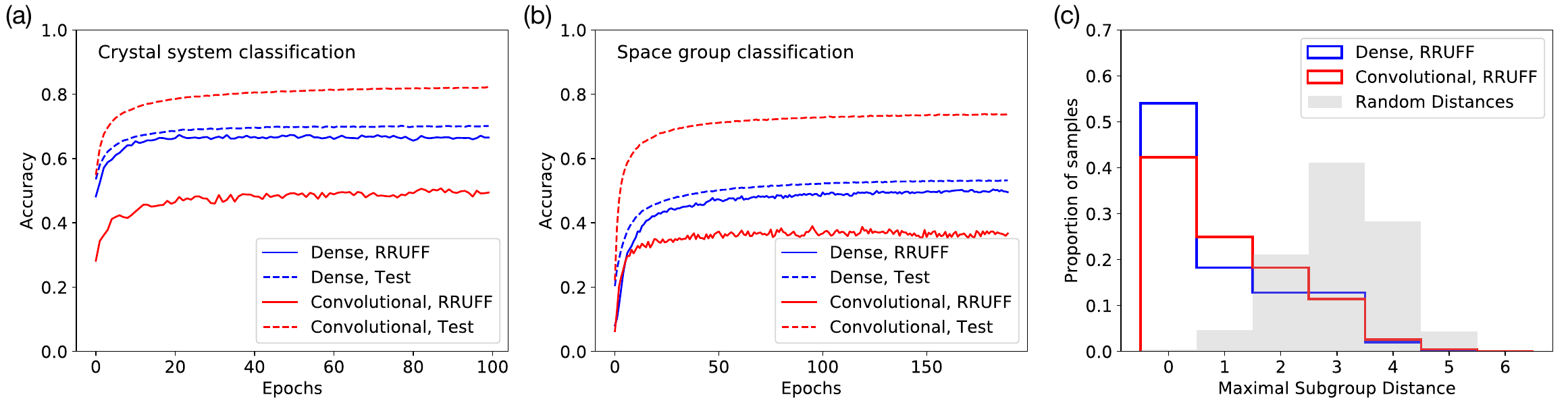}
    \caption{
    Network classification accuracies. (a) and (b) show the evolution of the network accuracies evaluated over the test set and the RRUFF database, which consists of experimentally obtained XRD patterns, for (a) the classification of crystal systems and (b) the classification of space groups. The accuracies are averaged over ten networks initialized with a different random seeds. While the convolutional network performs better on the training data in both cases, the dense network generalizes better to the experimental RRUFF database. (c) Histogram of the distances between network's prediction and correct space group classification. This distance is defined on the maximal subgroup graph.}
\label{fig: Network Training}
\end{figure*}

\section{Data preparation}
\label{sec: data}
The XRD patterns used for training the networks were computed theoretically from the crystal structure information obtained from the Inorganic Crystal Structure Database (ICSD)~\cite{bergerhoff1987crystallographic} which provides the crystallographic information file (CIF) of many inorganic substances. We removed duplicates (some crystal structures appeared more than once in the database) from the training data. With this restriction, a dataset of $128404$ samples was obtained.

The CIF contains the necessary information required to compute XRD patterns. Using the Python pymatgen library \cite{pymatgen}, we computed the theoretical XRD pattern in terms  of peak heights and positions from the CIF. The peak heights and positions are then convolved with a Gaussian function of variable standard deviation $\sigma$ to mimic the finite experimental resolution. 

However, these theoretically computed patterns still lack counting statistics and background signals present in real experimental data. A network trained using only such data may interpret the experimental noise as Bragg peaks potentially causing a misclassification. To simulate statistical noise, we augment our XRD patterns with a random signal drawn from a uniform distribution. In addition, we add a background signal $f(\theta)$ which is the sum of the following functions:
\begin{enumerate}
\item Smooth step functions:
\begin{equation}
\begin{split}
f_{\textrm{stepup}}(\theta)  &= h\left(\frac{1}{1+\textrm{exp}[a(\theta-\theta_{\textrm{step}})]}\right) \\
f_{\textrm{stepdown}}(\theta)  &= h\left( 1 - \frac{1}{1+\textrm{exp}[a(\theta-\theta_{\textrm{step}})]}\right)
\end{split}
\end{equation}
where $h$, $a$ and $\theta_{s}$ gives the height, steepness, and position of the step, respectively. 
\item Fourth-order polynomials 
\begin{equation}
f_{p}(\theta)  = \left| \sum_{n=0}^{4} a_{n}\theta^{n} \right|
\end{equation}
where the coefficient $a_n$ are chosen randomly.
\item To mimic enhanced background near the direct beam (small scattering angles 2$\theta$), but before the beam-stop, 
we used
\begin{equation}
f_{\textrm{bump}}(\theta) = h(n\theta)^{2}e^{-n\theta}
\end{equation}
where $h$ is setting the magnitude and $n$ is a measure of the steepness.
\end{enumerate}
The exact details of how these background and noise parameters are chosen is given in the appendix.

In Fig.~\ref{fig: XRD Sample}, we compare the XRD pattern produced using the above procedure with a true experimental data from the RRUFF database. The noise and background signals in the two patterns are relatively similar suggesting that the method we used to create the data could help the network generalize better to experimental conditions. As a side note, it can be seen that although both crystals in Fig.~\ref{fig: XRD Sample} correspond to space group 15, their XRD patterns barely have any similarities that can be easily picked out by the naked eye. This illustrates the significant difficulty in the classification problem that the network faces.

The full dataset is then divided into three sets: test, validation and training. The test and validation sets have a size of 7000 each, leaving us with 114404 samples in the training set.

\section{Results and Discussion}
\label{sec: results} 
\subsection{Training}
Using the dataset described in the previous section, we train the network architectures shown in Fig.~\ref{fig: Network Architecture}. The evolution of the classification accuracies of the networks over both the test set and the RRUFF database (real experimental data) during the training are shown in Fig.~\ref{fig: Network Training}. The accuracies were averaged over ten trained networks initialized with a different random seed. For the final prediction, the output of the network was averaged over the ensemble of ten networks. The final accuracies of the network ensembles are summarized in Table~\ref{tab: accuracies}. Figure~\ref{fig: Network Training}~(a) and (b) shows the results for the crystal systems and space groups, respectively.

\begin{table}[h]
\begin{tabular}{|c|c|c|c|c|}
\hline
\multirow{2}{*}{} & \multicolumn{2}{l|}{Crystal systems} & \multicolumn{2}{l|}{Space groups} \\ \cline{2-5} 
                  & Test set           & RRUFF            & Test set     & RRUFF           \\ \hline
Convolutional     & $85\%$           & $56\%$           &    $76\%$              &       $42\%$            \\ \hline
Dense             & $73\%$           & $70\%$           &         $57\%$         &    $54\%$      \\ \hline
\end{tabular}
\caption{Obtained classification accuracies of the convolutional and dense networks for the theoretically computed test set data and for the experimental RRUFF data.}
\label{tab: accuracies}
\end{table}

As expected, the networks are clearly more accurate in classifying the theoretical test data (which has the same noise and background structure as the training data) than the experimental data (which contains noise and background structures not present in the training data). This suggests that despite the addition of background and noise there are still important systematic differences between the theoretical and experimental data which may be one of the main obstacles for obtaining a higher classification accuracy on the experimental data. We also note that the correct classification may not be possible for instance in the case of non-centrosymmetric crystals in which the experiment averages over twin domains. However, it is remarkable that this discrepancy is much larger in the convolutional network as compared to the dense network. In fact this discrepancy is so large that, even though the former had a higher accuracy over the test set, the RRUFF database is significantly better classified by the dense network. This means that the dense network generalizes better to imperfect data.

\subsection{Classification Accuracy of Individual Space Groups}
\begin{figure}[h]
            \includegraphics[width=0.9\columnwidth]{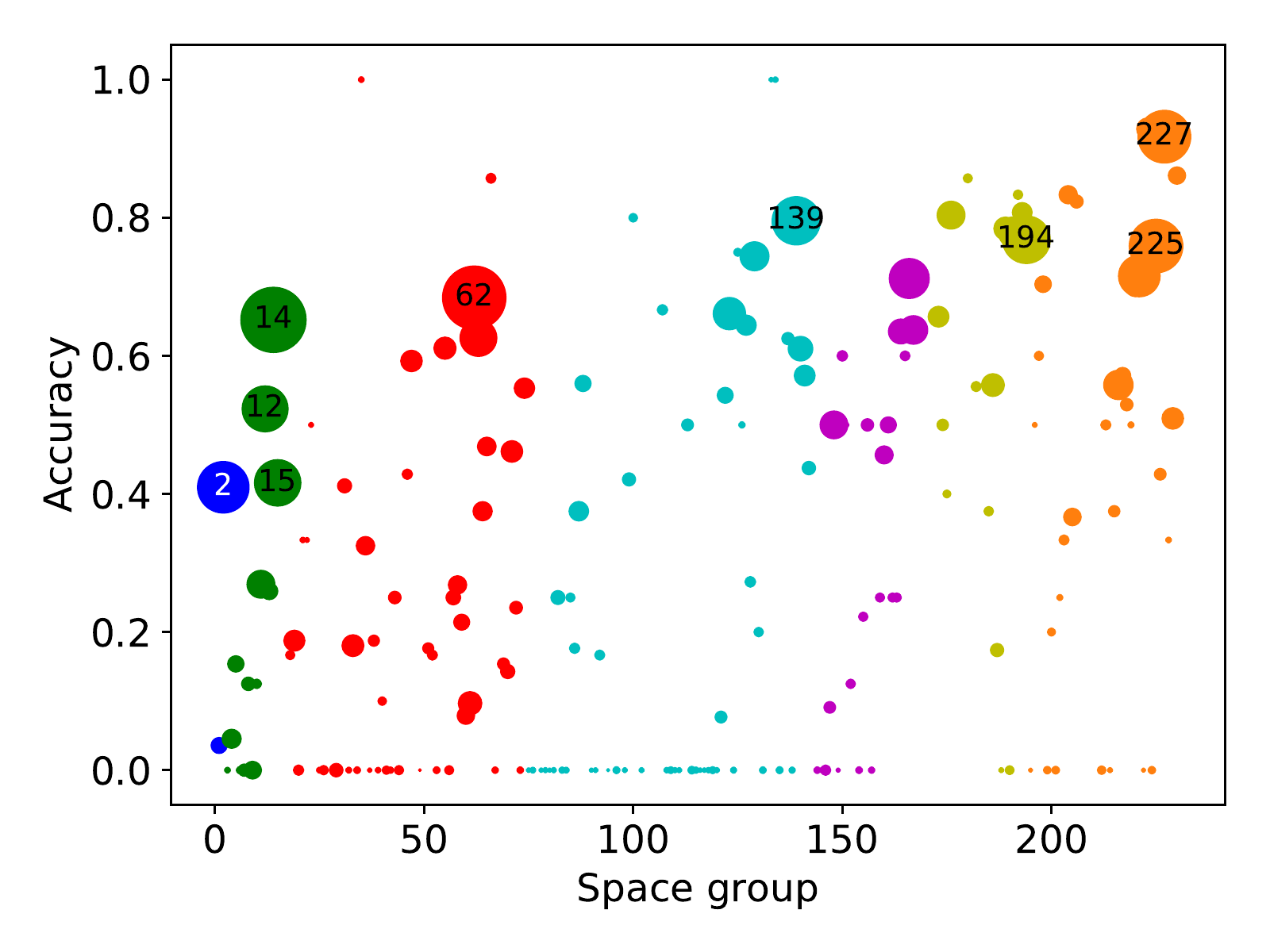}
    \caption{Classification accuracy of individual space groups. The colours denote the corresponding crystal system and the area of the circle indicates its relative abundance in the training set. Space groups corresponding to larger indices tend to contain more symmetry elements. For instance, space group 1 contains no point group symmetries whereas space group 230 contains the full set of cubic symmetries. Depicted here are the accuracies for the theoretically obtained test set, because the RRUFF database is too small to provide meaningful statistics for individual space groups. (There are around 800 samples for 230 space groups in the RRUFF database.)}
\label{fig: SG Distribution}
\end{figure}

Next, we look at the classification accuracies of individual space groups. Since the distribution of space groups within the training set is highly non-uniform, we expect that the trained network would perform better at classifying the more common space groups. As can be seen in Fig.~\ref{fig: SG Distribution}, this is indeed the case. Moreover, we also observed that the larger space groups (i.e., space groups with more symmetry elements) are more accurately classified. This is not surprising since with more symmetry elements there are more constraints on the structure factors giving rise to simpler XRD patterns with fewer Bragg peaks.

\subsection{Maximal Subgroup Distance}
To further elucidate the classification quality of the network, we consider the distance measure on the set of space groups to assess how far a wrong classification by the network is off the correct classification. This distance measure is defined through the concept of maximal subgroups. A maximal subgroup $\mathcal{B}$ of $\mathcal{A}$ is a proper subgroup such that no other proper subgroups $\mathcal{C}$ strictly contains $\mathcal{B}$. This allows us to construct a graph where the nodes represent the space group and two nodes are connected by an edge if one of the corresponding space groups is a maximal subgroup of the other. We can then define the maximal subgroup distance between two space groups to be the shortest path connecting their respective nodes. Intuitively, one may expect that $\mathcal{A}$ and $\mathcal{B}$ differ by a few symmetry elements such that their respective XRD patterns should be relatively similar and the network may confuse them more easily. Figure~\ref{fig: Network Training}~(b) shows the histogram of distances in the maximal subgroup graph for the network's prediction over both the test set and the RRUFF database. It clearly illustrates that even when the network's prediction is incorrect, it is far from random. It often lies closer to the correct prediction than one would expect from a random choice. This also suggest a possible alternative to interpret the classification output of the network: A space group prediction $\mathcal{X}$ indicates with high probability that the corresponding crystal is in $\mathcal{X}$ or in one of the neighbouring space groups in the maximal subgroup graph. 

\begin{figure*}[t]
            \includegraphics[width=1.0\textwidth]{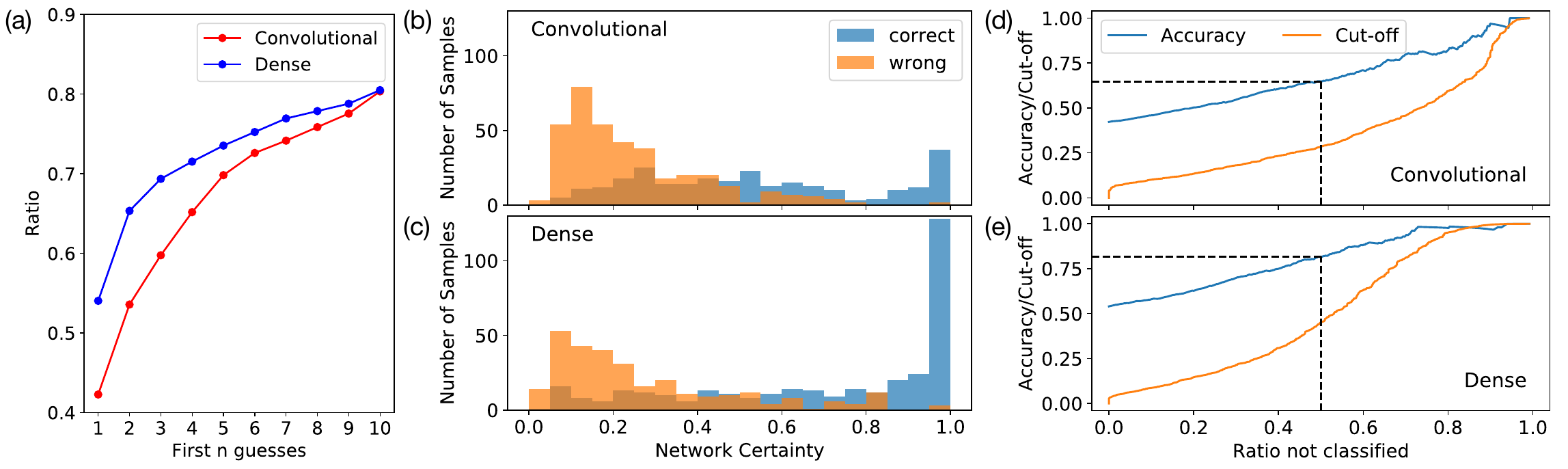}
    \caption{(a) Ratio of samples which are correctly classified by the $n$ highest outputs. (b) and (c) are histograms showing the number of samples in the RRUFF database which are classified correctly or wrongly with a given certainty of the network. (b) Convolutional network. (c) Dense network. (d) and (e) depicts how the accuracy (fraction of correctly classified samples on the RRUFF database) of the network changes as the certainty cut-off is varied. The orange line gives the ratio of unclassified samples as the cut-off is varied while the blue lines shows the accuracy of the network given a certain ratio of unclassified samples. The dotted lines indicate the accuracy when $50\%$ of the samples are unclassified. At this cut-off level, the accuracies are $65\%$ and $82\%$ for the (d) convolutional network and the (e) dense network, respectively.}
\label{fig: Certainty}
\end{figure*}

\subsection{Analysis of the Full Output Distribution of the Network}
So far, we have only considered the space group/crystal system that corresponds to the maximum of the output vector $\boldsymbol{y}$ as the classification result of the network. 
However, the network's output $\boldsymbol{y}$ is in fact a probability distribution over the possible categories that contains much more information. We define the network's first prediction (this is the same as the network's prediction result) as the most probable category, the second prediction as the next most probable and so on. 

In Fig.~\ref{fig: Certainty} (e), we show the first $10$ predictions of neural networks evaluated on the test set and RRUFF database. As expected, we can see from the decreasing distance between consecutive data points that the first prediction is more often correct than the second prediction, the second prediction is more often correct than the third and so on. This shows that there is additional useful information contained within the network's output on top of just its prediction which corresponds to its first prediction. Therefore, if we find that the network's prediction is wrong, the next most likely correct answer is indeed the second prediction. In other words, the probability distribution over the various categories as given by the network's output vector $\boldsymbol{y}$ is a good prior.

\subsection{Network Certainty}
Finally, we consider how to use network's certainty of its prediction to enhance the classification accuracies. Given a network output $\boldsymbol{y}$, the certainty is then defined by the probability of the network's prediction, i.e.,
\begin{equation}
\textrm{certainty} = \max_{i} \boldsymbol{y}_{i},
\end{equation}
where $i$ runs over all space groups/crystal systems. 

In Fig.~\ref{fig: Certainty}~(a) and (b), we show the classification certainties of both the convolutional and dense networks evaluated over the RRUFF dataset. The figures show that for the dense network, when the prediction is correct, the certainty is high. This result can be used to increase the accuracy of network classifications, if one allows the network to not classify inputs with low certainty, as follows: Let $\boldsymbol{y}$ be the network's output for some input XRD pattern that is to be classified. If the networks certainty is above a certain threshold, we accept the classification. Otherwise we say that the network is uncertain and leave the input unclassified. Figure~\ref{fig: Certainty}~(c) and~(d), shows the result of such a scheme: Placing the certainty threshold at $0.45$, we get a classification accuracy of around $82\%$ for the dense network on the RRUFF database at the cost of having half of the data unclassified. The rationale behind such a tradeoff is that it is better to know that the network is uncertain about a (possibly low-quality) input rather than for the network to give a wrong classification.

It is imperative to use Fig.~\ref{fig: Certainty} for a comparison between the convolutional and the dense network. Whereas the dense network's prediction is often correct when it is certain, the correlation between certainty and correct classification is much weaker in the convolutional network. For instance, if we impose the same certainty threshold of $0.45$ on the convolutional network, we would have around $69\%$ unclassified data, and if we impose a threshold such that only half of the data is unclassified, the final classification accuracy is only $65\%$. This once again suggests that the convolutional network is not generalizing well to the experimental data.

\section{Conclusion}
We have trained neural networks for the purpose of classifying powder XRD patterns by the respective crystal system and space group of the crystal. The training was performed with theoretically computed training data based on real crystal structures. In order for the trained network to classify experimentally obtained data, we added noise and artificially constructed background signals to the training data. 

In particular, we compared the performance of a convolutional network that was previously introduced in Ref.~\cite{ParkNNXRD2017} for the same purpose and a deep dense neural network. We find that although the convolutional network classifies theoretical data much more accurately, it generalizes poorly to experimental data. This is consistent with the results in Ref.~\cite{ParkNNXRD2017}, where it was found that even though the network had above $80\%$ accuracy on theoretical data, it failed on all of three experimental samples studied. We find that the deep dense network has a higher classification accuracy ($54\%$) on experimental data than the convolutional network ($42\%$). 

Next, to better understand the quality of the network's prediction, we used the concept of maximal subgroups to provide a distance measure on the set of space groups. We found that even though the network might give a wrong prediction, this predicted space group is not random but instead lies close to the correct answer.

Finally, we analyzed the network's certainty to enhance the classification accuracy. By allowing the network to be undecided about a subset of the predictions, we were able to enhance the classification accuracy of the dense network (on experimental data) to $82\%$ at the expense of leaving around half of the samples unclassified. This could be considered an improvement since it is helpful to know beforehand that a network's prediction is wrong. For the convolutional network, in contrast, we find that the certainty is not as good an indicator of correctness of the network's prediction.

Several routes for future improvements on our results present themselves and may lead to broad applications of machine learning techniques for crystallography. For one, the models we employ for artificial noise and background signals may be improved to increase the accuracy. Second, performant algorithms may be obtained by employing traditional and machine-learning algorithms, for instance to pre-process the training data. Finally, a growing experimental database of XRD samples may provide better, experimental training data.

\section*{Acknowledgments}
We would like to thank Ruggero Frison, Antonio Cervellino,  Denis Cheptiakov, Ekaterina Pomjakushina, Pascal Puphal, and Ann-Christin Dippel for discussions as well as comments on the manuscript, and for sharing of their diffraction data.
KC was supported by the European Unions Horizon 2020 research and innovation program (ERC-StG-Neupert-757867-PARATOP). JC acknowledge support from the Swiss National Science Foundation. The machine learning techniques were implemented using Keras~\cite{chollet2015keras} with a Tensorflow backend~\cite{tensorflow2015-whitepaper}.

\section*{Appendix: Data Preparation}

The XRD patterns used for training the networks were computed theoretically from the crystal structures information obtained from the Inorganic Crystal Structure Database \cite{bergerhoff1987crystallographic}. We denote the theoretically computed XRD pattern by 
$f_{\textrm{theory}}(\theta)$
which is defined on the domain $\theta_{\textrm{min}} < \theta < \theta_{\textrm{max}}$ and takes values in the codomain $0 \leq f_{\textrm{theory}}(\theta) \leq 1$ . However, these patterns still lack the noise and background signals present in real experimental data. A network trained using only such data may interpret the experimental noise as Bragg peaks potentially causing a misclassification. To introduce noise, we augment our XRD patterns with a random signal drawn from a uniform distribution $\mathcal{U}(0.002,0.02)$ between $0.002$ and $0.02$.

Next, to simulate a background we add a signal which is composed of four different functions. These four functions are selected with a $50\%$ probability each, i.e., not all four functions have to appear in our background signal. We also define a threshold $T$ given by
\begin{equation*}
T = \frac{0.1}{\textrm{number of different background functions chosen}},
\end{equation*}
which controls the amplitude of each background function. For convenience, we also define the relative angle $\theta_{\textrm{rel}} = \frac{\theta - \theta_{\textrm{min}}}{\theta_{\textrm{max}} - \theta_{\textrm{min}}}$.
The four background functions are given by:
\begin{enumerate}
\item Smooth step functions
\begin{equation}
\begin{split}
f_{\textrm{stepup}}(\theta)  &= h\left(\frac{1}{1+\textrm{exp}[a(\theta_{\textrm{rel}}-\theta_{\textrm{rel},\textrm{step}})]}\right), \\
f_{\textrm{stepdown}}(\theta)  &= h\left( 1 - \frac{1}{1+\textrm{exp}[a(\theta_{\textrm{rel}}-\theta_{\textrm{rel},\textrm{step}})]}\right),
\end{split}
\end{equation}
where $h$, $a$ and $\theta_{\textrm{rel},\textrm{step}}$ gives the height, steepness, and position of the step, respectively. These parameters are chosen as follows: $h$ is drawn a truncated normal distribution $h \in [0,T]$ with mean $\mu = T/3$ and standard deviation $\sigma = T/7$, $a$ is sampled from a uniform distribution $\mathcal{U}(10,60)$ and $\theta_{\textrm{rel},\textrm{step}}$ for the step-up (step-down) function is taken from a uniform distribution, with a width of $W = 1/7$, at the left edge (right edge) of the domain, i.e., $\mathcal{U}(0,W)$ for the step-up function and $\mathcal{U}(1- W,1)$ for the step-down function.
\item Polynomials (up to fourth order)
\begin{equation}
f_{p}(\theta)  = \left| \sum_{n=0}^{n_{\textrm{max}}} a_{n}\theta_{\textrm{rel}}^{n} \right|,
\end{equation}
where $n_{\textrm{max}}$ is a a random integer between $0$ and $4$. The coefficient $a_n$ are chosen according to:
\begin{equation}
a_n = \left\{
        \begin{array}{ll}
            0 & \quad \textrm{with probability } 0.5\\
            \frac{3T}{2(n_{\textrm{max}}+1)} \mathcal{U}(-1,1) & \quad \textrm{with probability } 0.5
        \end{array}
    \right.
\end{equation}
\item Bumps near the left edge of the XRD pattern (near $\theta_{\textrm{min}}$), to mimic the enhanced background near the direct beam
\begin{equation}
f_{\textrm{bump}}(\theta) = h(n\theta_{\textrm{rel}})^{2}e^{-n\theta_{\textrm{rel}}}
\end{equation}
where $h$ is the height of the bump and $n$ is a measure of the steepness. $h$ is drawn from a truncated normal distribution $h\in[0,3T/5]$ with mean $\mu = 2T/5$ and standard deviation $\sigma = 3T/35$ and $n$ is taken from a uniform distribution $\mathcal{U}(40,70)$.
\end{enumerate}

The full augmented XRD pattern is then given by
\begin{equation}
\begin{split}
f(\theta) =& f_{\textrm{theory}}(\theta) + f_{\textrm{stepup}}(\theta) + f_{\textrm{stepdown}}(\theta) \\
&+ f_{p}(\theta) + f_{\textrm{bump}}(\theta) \\
&+ \textrm{noise},\\
\end{split}
\end{equation}
where the parameters of the individual functions are chosen independently according to the procedure described above.

\bibliography{biblio}

\clearpage
\end{document}